%% file: dbs_isit07.tex
\documentclass[conference,twocolumn]{IEEEtran}
\pdfoutput=1

\newlength{\figwidth}
\usepackage[nosort]{cite}        
\usepackage{url} 
\usepackage{amsmath}
\usepackage{bbm}
\usepackage{graphicx}
\usepackage{paralist}
\usepackage[stretch=16,shrink=16,step=4]{microtype}
\usepackage{setspace}

\makeatletter
\def\@IEEEinterspaceratioM{0.265}
\def\@IEEEinterspaceMINratioM{0.1651}
\def\@IEEEinterspaceMAXratioM{0.38}

\def\@IEEEinterspaceratioB{0.31}
\def\@IEEEinterspaceMINratioB{0.19}
\def\@IEEEinterspaceMAXratioB{0.38}
\@IEEEtunefonts
\makeatother

\include{standard-macros}

\include{defs}

\newcommand{\pulsersii}{{{\sc Pulsers} Phase II}\xspace}

\begin{document}


\IEEEoverridecommandlockouts


\title{Capacity of Underspread Noncoherent WSSUS Fading Channels under Peak Signal Constraints}


\author{\authorblockN{Giuseppe Durisi, Helmut B\"{o}lcskei\\}\vspace{0.1\baselineskip}
\authorblockA{Communication  Technology Laboratory\\
 ETH Zurich, 8092 Zurich, Switzerland\\
E-mail: \{gdurisi,boelcskei\}@nari.ee.ethz.ch\\} \and
\authorblockN{Shlomo Shamai (Shitz)\\}\vspace{0.1\baselineskip}
\authorblockA{Technion, Israel Institute of Technology\\
 32000 Haifa, Israel\\
 E-mail: sshlomo@ee.technion.ac.il} \thanks{This work was supported in part by the Swiss
{\em Kommission f\"ur Technologie und Innovation (KTI)} under grant 6715.2 ENS-ES,
by the European Commission as part of the Integrated Project \pulsersii
under contract FP6-027142, and by the EC FP6 Network of Excellence NEWCOM.}}

\maketitle

\begin{abstract}
	We characterize the capacity of the general class of noncoherent  underspread wide-sense stationary uncorrelated scattering (WSSUS) time-frequency-selective Rayleigh fading channels, under peak constraints
	in time and frequency and in time only.
	Capacity upper and lower bounds are found which are explicit in the channel's scattering function and allow to identify the capacity-maximizing bandwidth for a given scattering function and a given peak-to-average power ratio.
\end{abstract}

\section{Introduction}\label{sec:introduction}

A well-known result in information theory states that, in the infinite-bandwidth limit, the capacity of a time-frequency (TF) selective, wide-sense stationary uncorrelated scattering (WSSUS)~\cite{bello63-12a} fading channel\footnote{Throughout the paper, whenever we speak of fading
channels, we shall refer to (without explicitly stating it) the
``noncoherent'' case where neither transmitter nor receiver have
access to channel state information (CSI) but both are aware of the
channel law.} equals the capacity of an additive white Gaussian noise (AWGN) channel with the same receive signal-to-noise ratio (SNR); capacity is achieved by codebooks that are ``peaky'' in both time and frequency~\cite{gallager68a,kennedy69}. 
It is also well-known that the AWGN channel capacity cannot be achieved, in the infinite-bandwidth limit, if a peak constraint is imposed~\cite{medard02-04a,subramanian02-04a}. In particular, as detailed below, different forms of 
peak constraints lead to different infinite-bandwidth capacity behavior.
\subsubsection*{Peak constraint in time}
A \closedform expression for the capacity of a WSSUS TF-selective fading channel under a peak constraint in time is not available in the literature. The rate achievable by frequency-shift-keying (FSK), in the infinite-bandwidth limit, has been shown by Viterbi~\cite{viterbi67-07a} to be given by
\bal
	 	\rateFSKinf=\underbrace{\intlim\frac{\Pave}{\noiseDen} \powerDopplerProfile(\doppler) \id \doppler}_{\cawgninf}
 	-\intlim \log\lefto(1+\frac{\Pave}{\noiseDen}\powerDopplerProfile(\doppler)\right)\!\id \doppler \label{eq: viterbi result}
\eal
where \Pave is the received power, \noiseDen stands for the one-sided noise spectral density and $\powerDopplerProfile(\doppler)$ is the power-Doppler profile~\cite{bello63-12a} of the TF-selective fading channel process. The infinite-bandwidth rate achieved by FSK constitutes a lower bound on the infinite-bandwidth capacity of the channel and equals the infinite-bandwidth capacity \cawgninf of an AWGN channel  (with the same receive SNR) minus a penalty term resulting from the peak constraint. In the absence of a peak constraint, the second term on the right-hand-side (RHS) of~\eqref{eq: viterbi result} can be made as small as desired by transmitting the FSK codewords with arbitrarily low duty cycle, and AWGN channel capacity can be achieved in the infinite-bandwidth limit~\cite[Sec. 8.6]{gallager68a}.

Expressions similar to~\eqref{eq: viterbi result} have been found in~\cite{sethuraman05-09a} for the capacity per unit energy and in~\cite{sethuraman06-07a,zhang07-01a} for the infinite-bandwidth capacity of time-selective, frequency-flat Rayleigh fading channels under a peak constraint in time.

\subsubsection*{Peak constraint in time and frequency}
From the results in~\cite{medard02-04a,subramanian02-04a} it follows that, under a peak constraint in time and frequency, the capacity of a TF-selective WSSUS fading channel goes to zero as the bandwidth approaches infinity.
An upper bound on the rate achievable by transmitting constant modulus symbols 
over a TF-selective WSSUS {\em underspread}~\cite{kozek97-03a} fading channel has been obtained in~\cite{schafhuber04}. This upper bound is explicit in the channel's scattering function and hints at the existence of a scattering-function-dependent, capacity-maximizing bandwidth.
\subsubsection*{Contributions}
We consider 
the general class of underspread~\cite{kozek97-03a} continuous-time
WSSUS TF-selective Rayleigh fading channels under peak constraints 
\begin{inparaenum}[(i)]
	\item in time and 
	\item in time and frequency.\footnote{The peak constraints, made precise in Section~\ref{sec:model}, will be imposed in the channel's eigenspace rather than on the continuous-time transmit signal.}
\end{inparaenum}
Our specific contributions are  summarized as follows:
\begin{itemize}
 \item  {\em Peak constraint in time:} We recover Viterbi's infinite-bandwidth lower bound~\eqref{eq: viterbi result} using an alternative proof.
 \item  {\em Peak constraint in time and frequency:} We generalize the results in~\cite{schafhuber04}, which dealt exclusively with constant modulus inputs, by providing upper and lower bounds on capacity. Both bounds are explicit in the channel's scattering function and reveal the existence of an optimum, capacity-maximizing bandwidth, which depends on the scattering function and the peak-to-average power ratio.
\end{itemize}
\subsubsection*{Proof techniques}
Our entire analysis is built on the fact that underspread channels have a well-structured set of TF-localized approximate eigenfunctions~\cite{kozek97-03a}.  The main proof techniques used in the paper are based on the relation between mutual information and minimum mean-square error (MMSE) discovered in~\cite{guo05-04a}, on a generalization of Szeg\"{o}'s theorem on the asymptotic eigenvalue distribution of Toeplitz matrices~\cite{szego} to the case of block-Toeplitz matrices with Toeplitz blocks\footnote{In the remainder of the paper,  block-Toeplitz matrices with Toeplitz blocks will simply be referred to as block-Toeplitz matrices.}~\cite{tyrtyshnikov98}, and on a property of
the information divergence of orthogonal signaling schemes first
presented in a different context in~\cite{butman73}. 

\subsubsection*{Notation}
Uppercase boldface letters denote matrices and  lowercase boldface letters designate vectors. The superscripts $\tp{}$, $\herm{}$ and $\conj{}$ stand for transpose, conjugate transpose and element-wise conjugation, respectively. $\determinant{\XX}$  denotes the determinant  of the matrix \XX, \I stands for the identity matrix of appropriate size, and $\diag{\X}$  denotes a diagonal square matrix having the elements of the vector \X on its main diagonal. $\EV{}{\cdot}$ is the expectation operator. $\delta(\time)$ stands for the Dirac delta function, and  $\delta_{i,j}=1$ if $i=j$ and $0$ else. All logarithms are to the base $\nepero$. If a random variable (RV) \inpdiscr has distribution $\probinpdiscr$, we write $\inpdiscr\distributedas \probinpdiscr$. Finally,  $\circnorm(0,\sigma^{2})$ denotes a circularly symmetric complex Gaussian RV with variance $\sigma^{2}$.

\section{Channel and System Model}\label{sec:model}

We consider an ergodic WSSUS TF-selective Rayleigh  fading channel \opH  with input-output relation
\bal\label{eq: input-output relation}
	\outpnonoise(\time) = (\opH \inp)(\time) = \int_{\delay}\!\ch{\time,\time-\delay} \inp(\time-\delay) \id \delay
\eal
where the impulse response $\ch{\time,\time'}$ is a two-dimensional zero-mean complex Gaussian random process. The time-varying transfer function of the channel is defined as~\cite{bello63-12a}
\ba
	\trafun{\time,\freq} =\int_{\delay} \!\ch{\time,\time-\delay} e^{-j 2 \pi \freq \delay} \id\delay.
\ea
The  channel's scattering function $\scafun{\delay,\doppler}$~\cite{kennedy69} is given by 
\ba
	\scafun{\delay,\doppler}=\int_{\time}\int_{\freq}\corfun{\time,\freq}\esp{-}{(\doppler\time-\delay\!\freq)} \id \time \id \freq
\ea
where 
%
%
%
$	\corfun{\time,\freq}=\EV{}{ \trafun{\time+\time',\freq+\freq'} {\trafunconj{ \time',\freq'} } }$
%
%
%
denotes the channel's TF-correlation function.
We invoke the common assumption~\cite{kozek97-03a} of a scattering function that is compactly supported within the rectangle $[-\maxDelay,\maxDelay] \times [-\maxDoppler, \maxDoppler]$, i.e., 
$\scafun{\delay,\doppler} = 0$ for $(\delay, \doppler) \notin  [-\maxDelay, \maxDelay] \times [-\maxDoppler,\maxDoppler]$.
Defining the spread of the channel as the area of this
rectangle, $\spread=4\maxDelay \maxDoppler$, the channel is said
to be underspread if $\spread \leq 1$ and overspread if
$\spread>1$~\cite{kozek97-03a}. The
underspread assumption is relevant as most mobile radio channels are
highly underspread.

The key idea turning~\eqref{eq: input-output relation} into a discrete problem is to recognize that underspread channels are approximately diagonalized by orthonormal Weyl-Heisenberg bases~\cite{kozek97-03a}, which are obtained
by TF-shifting of a normalized function $g(\time)$ according to
$g_{\dfdt}(\time) = g(\time - \dtime\tstep ) \esp{}{\dfreq \fstep \time}$, where the grid parameters $\tstep$
and $\fstep$ have to satisfy $\tstep \fstep \geq 1$. For $\tstep \leq 1/(2\maxDoppler)$ and $F \leq 1/(2\maxDelay)$, and hence $\TF \leq 1/\spread$, the impulse response $\ch{\time,\time'}$ of an underspread fading channel satisfies~\cite{durisi07a}
\bal
	  \ch{\time,\time'}\approx\sum_{\dtime=-\infty}^{\infty}\sum_{\dfreq=-\infty}^{\infty}  \trafun{\sampleTsampleF} g_{\dfdt}(\time)
    \conj{g_{\dfdt}}(\time')\label{eq: approximate diagonalization}.
\eal
The scalar channel coefficients  \pagebreak
$\trafun{\sampleTsampleF}$  are circularly symmetric complex Gaussian 
RVs with zero mean, variance%
$\sigmaH=\int_{\delay}\int_{\doppler} \scaFtn \id\delay \id\doppler$, 
and correlation function   
\ba
	\corfun{\dtime\tstep,\dfreq\fstep}=\EV{}{\trafun{(\dtime+\dtime')\tstep,(\dfreq+\dfreq')\fstep} \trafunconj{\sampleTsampleFprime}}.
\ea %

Motivated by the (approximate)
diagonalization~\eqref{eq: approximate diagonalization}, we write the transmit signal $\inp(\time)$ as 
\bal\label{eq: input signal}
    \inp(\time)=\sum_{\dtime=-\infty}^{\infty} \sum_{\dfreq=0}^{\fslots-1} \inpdiscr_{\dfdt}g_{\dfdt}(\time)
\eal
where the $\inpdiscr_{\dfdt}$ are the information-bearing data symbols. This
modulation scheme corresponds to pulse-shaped orthogonal frequency-division multiplexing (OFDM) 
with OFDM symbol duration $\tstep$ and
subcarrier spacing $\fstep$. The transmit-signal bandwidth is given by $\bandwidth= \fslots\fstep$. 
With the received signal
$\outp(\time) = \outpnonoise(\time)+\noise(\time)$
and 
$\noise(\time)$
circularly symmetric additive white Gaussian noise so that\footnote{In the remainder of the paper (apart from Section~\ref{Sec: numerical evaluation}), we normalize the one-sided
noise spectral density according to $\noiseDen=1$.}
$\EV{}{z(t)z^{*}(t')} = \delta(t - t')$,
 the receiver computes the
inner products $\outdiscr_{\dfdt} = \langle \outp, g_{\dfdt}\rangle$.
Exploiting the orthonormality of the
basis functions $g_{\dfdt}(\time)$, we obtain the overall input-output
relation
\bal
    \outdiscr_{\dfdt} = \chdiscr_{\dfdt} \inpdiscr_{\dfdt} + \noisediscr_{\dfdt} \label{generalch}
\eal
where $\chdiscr_{\dfdt}=\trafun{\sampleTsampleF}$ and $\media\{\noisediscr_{\dfdt}\conj{\noisediscr}_{\dfdtprime}\} =
\delta_{\dtime,\dtime'}\delta_{\dfreq,\dfreq'}$. 
In summary, we transmit and receive on the channel's (approximate)
eigenfunctions (which are TF-translates of the prototype function $g(\time)$), thereby realizing a tiling of the TF-plane. For a detailed discussion of the approximation~\eqref{eq: approximate diagonalization} and the consequences of restricting the class of input signals to~\eqref{eq: input signal}, the interested reader is referred to~\cite{durisi07a}. In the following, we refer to the index \dtime as representing the ``time-domain'' and the index~\dfreq as representing the ``frequency-domain''.
%

We assume that each channel use takes place over \fslots subcarriers and \tslots
OFDM symbols. 
 The \fslots-dimensional  vector containing the data symbols transmitted in the $\dtime$th OFDM symbol ($\all{\dtime}{\tslots}$) is denoted as 
$\X_{\dtime}=\tp{[\inpdiscr_{\dtime,0}\:\: \inpdiscr_{\dtime,1}\:\:\cdots\:\: \inpdiscr_{\dtime, \fslots-1}]}.$
The~vectors $\Y_{\dtime}$, $\Hv_{\dtime}$, and $\Z_{\dtime}$ are defined correspondingly. 
Furthermore, we define  the $\tslots\fslots$-dimensional vector containing the data symbols transmitted in one channel use as 
 $\X=\tp{[\tp{\X_0}\:\: \tp{\X_1} \:\: \cdots\:\: \tp{\X_{\tslots-1}}]}.$
Again, $\Y$, $\Hv$ and $\Z$ are defined correspondingly. Finally, we denote the covariance matrix of the channel vector \Hv
by $\cov{\Hv}=\EV{}{\Hv\herm{\Hv}}$. Since $\chdiscr_{\dfdt}$ is WSS in \dtime and \dfreq, the matrix $\cov{\Hv}$ is block-Toeplitz.

The input-output relation corresponding to one channel use can now be written as
\ba
 	\Y=\diag{\Hv} \X + \Z.
\ea
Note that the channel in~\eqref{generalch} will in general not be memoryless because the channel gains $\chdiscr_{\dfdt}$ are correlated across time index \dtime and across frequency index \dfreq.
Throughout the paper, we assume an average-power constraint  according to $\avP$. 
In addition, setting $\Ppeak=\ppar \Pave$, with $\ppar\geq 1$, we impose a peak constraint distinguishing the following two cases:
\begin{enumerate}[i)]
\item {\em Peak constraint in time and frequency:} \pagebreak The data symbols satisfy 
 \bal\label{def: PeakPerTF}
 	|\inpdiscr_{\dfdt}|^2 \leq {\Ppeak \tstep}/{\fslots}\qquad  \text{a.s.}\:\: \forall \dfdt
\eal
where a.s. stands for \textit{almost surely}. 
\item {\em Peak constraint in time:}  The data symbols satisfy  
 \bal\label{def: PeakPerSymb}
 	\|\X_{\dtime}\|^2 \leq \Ppeak \tstep \qquad \text{a.s.} \:\: \forall \dtime.
\eal
\end{enumerate} 
While Condition i) prohibits 
peakiness in time {\em as well as} peakiness in frequency, Condition ii) prohibits peakiness in time only, still allowing for a signaling scheme that is peaky in frequency.
Note that we impose peak constraints in the channel's eigenspace (i.e., on the data symbols $\inpdiscr_{\dfdt}$) rather than on the continuous-time transmit signal $\inp(\time)$.
\section{Capacity Bounds under Peak Constraints in Time and Frequency}
For a given bandwidth $\bandwidth=\fslots \fstep$, the capacity (in nat/s) of the channel~\eqref{generalch}
is defined as
\begin{align}\label{def: capacityPeakTF}
 	C(\bandwidth)=\limK \frac{1}{\tslots \tstep} \supPeakAve I(\Y;\X)
\end{align}
where the supremum is taken over the set of input distributions $\probx$  satisfying the average-power constraint $\avP$ and the peak constraint~\eqref{def: PeakPerTF}.
We shall next derive two upper bounds and a lower bound 
on~\eqref{def: capacityPeakTF}.
The first upper bound and the lower bound generalize the bounds for the frequency-flat, time-selective fading
case reported in~\cite[Prop. 3.1]{sethuraman06-07a}
and~\cite[Prop. 2.2]{sethuraman05-07a}, respectively, to the TF-selective underspread
case. 
In contrast to~\cite{sethuraman06-07a} and~\cite{sethuraman05-07a}, our proof
techniques do not explicitly rely on the relationship between the lag-one mean-square prediction error
of a WSS process and its spectral measure~\cite[Th. 4.3]{doob}. 
Instead, we use the relation between mutual information and MMSE discovered in~\cite{guo05-04a}, and a generalization of Szeg\"{o}'s theorem on the asymptotic eigenvalue distribution of Toeplitz matrices  to the case of block-Toeplitz matrices~\cite{tyrtyshnikov98}. 
The second upper bound is standard and is obtained by assuming perfect CSI at the receiver.
\subsection{First Upper Bound}
%
%
\begin{Theor}\label{Theorem: ub1}
The capacity of a TF-selective underspread Rayleigh fading channel, with scattering function $\scaFtn$, under the average-power constraint \avP and the peak constraint~\eqref{def: PeakPerTF} with $\Ppeak=\ppar \Pave$, is upper-bounded as $C(\bandwidth)\leq \UBone(\bandwidth)$, where
\bal
 \UBone(\bandwidth)&=     \frac{\bandwidth}{\TF} \log\lefto( 1+\alphanew  \Pave\frac{\TF}{\bandwidth}\sigmaH \right)-  \alphanew A(\bandwidth,\ppar) \label{def: UB1 cont} 
\eal
with
\bal
 	\alphanew&=\min\lefto\{ 1, \frac{\bandwidth}{\TF}\left(\frac{1}{A(\bandwidth,\ppar)} -\frac{1}{\Pave\sigmaH}\right) \right\}\label{eq: min for therorem UB1} 
\eal
and
\ba
 	A(\bandwidth,\ppar)&=\frac{\bandwidth}{\ppar}\inttaunu{1+\frac{\ppar \Pave}{\bandwidth}\scaFtn} \notag.
\ea
\end{Theor}
\begin{proof}
A sketch of the proof  can be found in Appendix~\ref{App: proof of theorem 1}.
\end{proof}
The upper bound~\eqref{def: UB1 cont} generalizes the upper bound~\cite[Eq. (2)]{schafhuber04}, which holds only for constant modulus signals, i.e., for $\abs{\inpdiscr_{\dfdt}}=\mathrm{const.},$ $\forall{\dfdt}$. The bounds~\eqref{def: UB1 cont} and~\cite[Eq. (2)]{schafhuber04} are both explicit in the channel's scattering function, have similar structure and coincide for $\ppar=1$ when $\alphanew=1$ in~\eqref{eq: min for therorem UB1}.

\subsection{Perfect Receive CSI Upper Bound}

A straightforward upper bound on~\eqref{def: capacityPeakTF} is obtained by assuming perfect CSI at the receiver and neglecting the peak constraint~\eqref{def: PeakPerTF}, but retaining the average-power constraint. 
The resulting bound is given by  
%
%
%
%
%
\bal\label{def: UB2 cont}
 	\UBtwo(\bandwidth)=  \frac{\bandwidth}{\TF}\: \EV{\chdiscr}{\log\lefto(1 + \frac{\Pave\, \TF}{\bandwidth} |\chdiscr|^2 \right)}
\eal
with $\chdiscr\distributedas\circnorm(0,\sigmaH)$.
\subsection{Lower Bound}

%
%

\begin{Theor}\label{Theorem: lower bound} 
	The capacity of a TF-selective underspread Rayleigh fading channel, with scattering function $\scaFtn$, under the average-power constraint \avP and the peak constraint~\eqref{def: PeakPerTF} with $\Ppeak =\ppar \Pave$, is lower-bounded, for large enough $\bandwidth$\!, as $C(\bandwidth)\geq \LB(\bandwidth)=\max_{1 \leq \gamma \leq \ppar} \LB(\bandwidth,\gamma)$, where
\begin{multline} 
	 \LB(\bandwidth,\gamma)=      \frac{\bandwidth}{\gamma\TF} I(\outdiscr;\inpdiscr|\chdiscr) \\ 
	 - \frac{\bandwidth}{\gamma}\inttaunu{1+\frac{\gamma\Pave}{\bandwidth}\scaFtn}\label{def: LB cont2}.
\end{multline}
The first term on the RHS of~\eqref{def: LB cont2} is the perfect-CSI mutual information of a scalar channel with input-output relation $\outdiscr=\chdiscr\inpdiscr+\noisediscr$, where $\chdiscr\distributedas \circnorm(0,\sigmaH)$, $\abs{\inpdiscr}^{2}=\gamma\Pave\tstep/{\fslots}$ a.s., and $\noisediscr\distributedas\circnorm(0,1)$.
\end{Theor}
\begin{proof}
	A sketch of the proof  can be found in Appendix~\ref{App: proof of theorem 2}.
\end{proof}
Noting that constant modulus constellations are second-order optimal in the low-SNR regime (see~\cite[Th. 14]{verdu02-06a}), we obtain the following explicit expression for~\eqref{def: LB cont2}: 
\begin{multline}
 	\LB(\bandwidth,\gamma)  \approx  \Pave \sigmaH - \frac{\gamma ( \Pave \sigmaH)^2 \TF}{\bandwidth}  \\
	- \frac{\bandwidth}{\gamma}\inttaunu{1+\frac{\gamma\Pave}{\bandwidth}\scaFtn}\label{def: LB approx beta}.
\end{multline}
\subsection{Discussion of the Bounds}

The upper bound~\eqref{def: UB1 cont} and the lower bound~\eqref{def: LB cont2} have the same structure in the sense of being the difference of the mutual information of a memoryless channel and a penalty term that depends on the fading channel memory. Moreover, both bounds are explicit in the channel's scattering function. 
Unlike the upper bound~\eqref{def: UB1 cont}, the lower bound~\eqref{def: LB cont2} holds only for large bandwidths as the penalty term on the RHS of~\eqref{def: LB cont2} admits a closed-form (integral) expression only for sufficiently large \bandwidth. 
\subsection{Numerical Evaluation of the Bounds}\label{Sec: numerical evaluation}
\begin{figure}[t] \centering
    \includegraphics[width=0.46\textwidth]{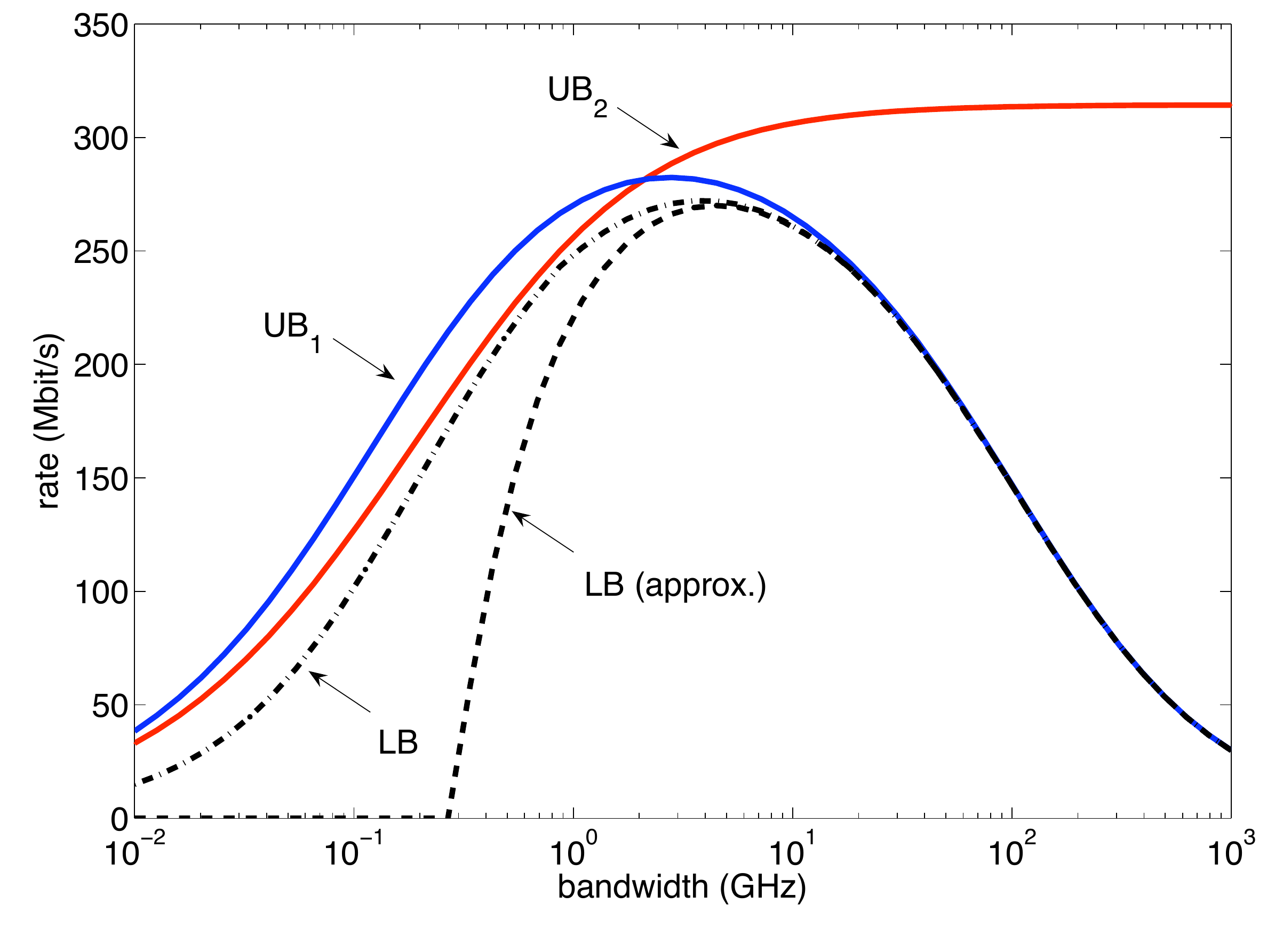}
  \caption{\UBone, \UBtwo, \LB, and the approximate lower bound~\eqref{def: LB approx beta} for $\ppar=1$.}\label{Fig: bounds TF-peak}
\end{figure}
In this section, we provide  numerical results based on 
the upper bounds~\eqref{def: UB1 cont} and~\eqref{def: UB2 cont}, and on the lower bound~\eqref{def: LB cont2},   where the first term in~\eqref{def: LB cont2} is evaluated \pagebreak numerically for QPSK modulation    using the algorithm proposed in~\cite{he05-05a}. 
We use IEEE 802.11a related
system parameters, i.e., $\TF=1.25$, transmit power equal to $1$~mW and one-sided
noise spectral density $\noiseDen=4.14 \cdot 10^{-21}$~W/Hz.  
The channel's scattering function is assumed to be brick-shaped with  $\spread=10^{-3}$ (i.e., the channel is highly underspread) and path loss $\sigmaH=-90$ dB.
Fig.~\ref{Fig: bounds TF-peak}  shows $\UBone(\bandwidth)$, $\UBtwo(\bandwidth)$,  
$\LB(\bandwidth)$ and the approximation~\eqref{def: LB approx beta}
for  $\ppar=1$. 
We can see    
that \UBone and \LB take on a maximum at the {\em critical} bandwidth of approximately $1$ GHz and then approach
zero as the bandwidth increases.
The approximation~\eqref{def: LB approx beta} is accurate for bandwidths above the critical bandwidth, and very loose otherwise. 
The effect of the capacity decreasing for bandwidths exceeding the critical bandwidth, and eventually going to zero, is known in the literature as {\em overspreading}~\cite{medard02-04a,subramanian02-04a,telatar00-07a}.
Overspreading occurs since the peak constraint~\eqref{def: PeakPerTF} prohibits peakiness of the signaling scheme in time as well as in frequency.

%
\section{Infinite-Bandwidth Capacity under Peak Constraint in Time}

We shall next relax the peak constraint in time and frequency and impose a peak constraint in time only according to~\eqref{def: PeakPerSymb}, while maintaining the average-power constraint \avP. In addition, we focus on the infinite-bandwidth limit only. A capacity lower bound, which is explicit in the channel's scattering function, will allow us to conclude that the overspreading phenomenon discussed in the previous section can be eliminated by allowing the transmit signal to be peaky in frequency. 
We start by defining the infinite-bandwidth capacity of the channel~\eqref{generalch} as
\ba
 	\infcapacity=\limK  \limN \supPeakSymbAve \frac{1}{\tslots \tstep} I(\Y; \X)
\ea
where the supremum is taken over the set of input distributions $\probx$ satisfying the average-power constraint $\avP$ and the peak constraint~\eqref{def: PeakPerSymb}. 
We derive a lower bound on \infcapacity by evaluating ${1}/{(\tslots \tstep)} \cdot I(\Y; \X)$ in the limit $\tslots,\fslots\to \infty$ for a specific signaling scheme. This signaling scheme, described in detail in~\cite{durisi07a} mimics the FSK scheme used in~\cite{viterbi67-07a}, and can be interpreted as a generalization (to channels with memory in time and frequency) of the on-off FSK scheme proposed in~\cite{gursoy06} for memoryless (Ricean) fading channels. The resulting lower bound is summarized as follows.
\begin{Theor}\label{Theor: infinite bandwidth} 
	The infinite-bandwidth capacity of a TF-selective 
	underspread Rayleigh fading channel, with scattering function  $\scaFtn$, under the average-power constraint \avP  and the peak constraint~\eqref{def: PeakPerSymb} with $\Ppeak=\ppar \Pave$, 
	 is lower-bounded 
	as $\infcapacity \geq \LB(\ppar)$, where 
 	\bal
 	 	 	\LB(\ppar) &= \Pave\sigmaH -\frac{1}{\ppar} \intnu{1 + \ppar \Pave  \scafuntilde{\doppler}} 
 	 	 	\label{def :Viterbi} 
	\eal
	with
 	 $	\scafuntilde{\doppler} =\int_{\delay} \scaFtn \id\delay $
 	denoting the power-Doppler profile~\cite{bello63-12a} of the channel.
\end{Theor}
\begin{proof}
	The proof of this result can be found in~\cite{durisi07a}.
\end{proof}

The lower bound in~\eqref{def :Viterbi} coincides with Viterbi's lower bound~\eqref{eq: viterbi result} when $\ppar=1$. The proof technique used to obtain Theorem~\ref{Theor: infinite bandwidth} is, however, conceptually different from that in~\cite{viterbi67-07a}. Specifically,  in~\cite{viterbi67-07a} an appropriate choice of the codebook, namely FSK, reduces a WSSUS TF-selective fading channel with scattering function \scaFtn to an effective frequency-flat, time-selective fading channel with power-Doppler profile $\scafuntilde{\doppler}$. A Karhunen-Lo\`{e}ve decomposition of the resulting effective time-selective fading channel then leads to a \closedform expression for the error exponent, which finally yields the capacity expression~\eqref{eq: viterbi result}. The proof of Theorem~\ref{Theor: infinite bandwidth}, on the other hand, starts from an eigenfunction decomposition of the WSSUS TF-selective channel's impulse response $\ch{\time,\time'}$, and establishes an infinite-bandwidth capacity lower bound by computing the rate achievable for a specific signaling scheme (which mimics FSK). The main tool used in the proof of Theorem~\ref{Theor: infinite bandwidth} is a property of the information divergence of FSK constellations, first presented in a different context in~\cite{butman73}.
\appendices

\section{}\label{App: proof of theorem 1}

\paragraph*{Sketch of the proof of Theorem~\ref{Theorem: ub1}}
Fix $\tslots$ and $\fslots$. Following~\cite{sethuraman06-07a}, we use the chain rule for mutual information to write the supremum in~\eqref{def: capacityPeakTF} as
\begin{multline}\label{eq: App 1 chain rule}
	\supPeakAve I(\Y;\X) = \supPeakAve \lefto\{ I(\Y;\X,\Hv) - I(\Y;\Hv|\X) \right \}  \\
	\leq  \supalpha \Biggl\{  \supPeakRho  I(\Y;\X,\Hv) - \infPeakRho I(\Y;\Hv|\X)\Biggr\}
\end{multline}
where~\eqref{eq: App 1 chain rule} follows by rewriting the supremum over $\probx$  as  a double supremum over $\alpha \in [0, 1]$ and over the set of input distributions  $\probprime$ satisfying \avalpha and the peak constraint~\eqref{def: PeakPerTF}.
Define $\XX=\diag{\X}$. The first term on the RHS of~\eqref{eq: App 1 chain rule} can be upper-bounded as follows
\vspace{-1mm}
\bal
	&\supPeakRho  I(\Y;\X,\Hv) \stackrel{(a)}{\leq}  \sup_{\avalpha} \logdet{ \I+ \EV{}{ \XX \cov{\Hv}  \herm{\XX}}}\notag \\
	    &\quad \qquad\stackrel{(b)}{\leq} \sup_{\avalpha}  \sum_{\dtime=0}^{\tslots-1}\sum_{\dfreq=0}^{\fslots-1} \log\lefto(1 + \EV{}{\abs{\inpdiscr_{\dfdt}}^{2}}\sigmaH\right)\notag \\
	&\quad \qquad \stackrel{(c)}{\leq} \tslots\fslots  \log\lefto( 1 + \frac{\alpha\Pave\tstep  }{\fslots}\sigmaH \right)\label{eq: App 1 concavity}
\eal
where~(a) follows by ignoring the peak constraint 
and upper-bounding $I(\Y;\X,\Hv)$
 by the capacity of an AWGN channel with input $\XX\Hv$,~(b) follows from Hadamard's inequality, and~(c) follows from the concavity of the $\log$ function. 
The second term on the RHS of~\eqref{eq: App 1 chain rule} can be lower-bounded as
\bal\label{eq: lower bound-second term}
	\infPeakRho I(\Y;\Hv|\X) \geq c \inttaunulog{1 + \frac{\ppar \Pave}{\fslots \fstep}\scaFtn}
 \eal
with $c={\alpha\tslots\fslots\tstep\fstep}/{\ppar}$. The proof of~\eqref{eq: lower bound-second term}, detailed in~\cite{durisi07a}, is based on the relation between mutual information and MMSE discovered in~\cite{guo05-04a} and on the closed-form expression for the noncausal MMSE of a two-dimensional stationary random process obtained in~\cite{helstrom67}.
Inserting~\eqref{eq: App 1 concavity} and~\eqref{eq: lower bound-second term} in~\eqref{eq: App 1 chain rule}, dividing by $\tslots\tstep$, and using $\bandwidth=\fslots\fstep$, we get
\begin{multline}
	\UBone(\bandwidth) = \supalpha \Biggl \{ \frac{\bandwidth}{\tstep\fstep} \log\lefto( 1 + \frac{\alpha \Pave \tstep\fstep}{\bandwidth}\sigmaH \right) \\
- \frac{\alpha\bandwidth}{\ppar} \inttaunulog{ 1 +\frac{\ppar \Pave}{\bandwidth}\scaFtn} \Biggr \}. \notag
\end{multline}
As the function the supremum is taken over is concave in $\alpha$, the maximizing value given in~\eqref{eq: min for therorem UB1} is unique and can be obtained by differentiating with respect to $\alpha$. 

\section{}\label{App: proof of theorem 2}

\paragraph*{Sketch of the proof of Theorem~\ref{Theorem: lower bound}}
For the sake of simplicity of exposition, we outline the proof for $\ppar=1$ only. The result for general \ppar follows from a simple time-sharing argument~\cite[Corollary 2.1]{sethuraman05-07a}. We first lower-bound $C(\bandwidth)$ in~\eqref{def: capacityPeakTF} by assuming a specific input distribution, namely, by letting \X have independent and identically distributed (\iid), constant modulus entries, each of which satisfies $\abs{\inpdetscal}^2=\Pave \tstep/\fslots$. In the following, vectors $\X$ with this property will be denoted as \Xiid.  Next, we use the well-known inequality 
\bal
		I(\Y;\Xiid) &= I(\Y;\Xiid,\Hv)- I(\Y;\Hv|\Xiid) \notag \\
		                        &\geq I(\Y;\Xiid|\Hv)  - I(\Y;\Hv|\Xiid).\label{eq: App II  mut inf lb}
\eal
The \iid and the constant modulus assumptions imply that 
\bal
	I(\Y;\Xiid|\Hv)&= \tslots\fslots I(\outdiscr;\inpdiscriid|\chdiscr) \label{eq: App II coherent part} 
\eal
and %
\bal
	I(\Y;\Hv|\Xiid)&=\logdet{\I + \frac{\Pave \tstep}{\fslots} \cov{\Hv}} \label{eq: App II constant modulus}.
\eal
Next, by inserting~\eqref{eq: App II coherent part} and~\eqref{eq: App II constant modulus} in~\eqref{eq: App II  mut inf lb}, dividing by $\tslots \tstep$, and  taking the limit  $\tslots \to \infty$, we obtain 
\begin{multline}\label{eq: App II lb discrete time}
	C(\fslots \fstep)\geq \LB(\fslots \fstep)=\frac{\fslots}{\tstep} I(\outdiscr;\inpdiscriid|\chdiscr) \\
	-\limK \frac{1}{\tslots\tstep} \logdet{\I + \frac{\Pave\tstep}{\fslots} \cov{\Hv}}.
\end{multline}
 Finally, we use the generalization of Szeg\"o's theorem to block-Toeplitz matrices  provided in~\cite{tyrtyshnikov98} to show that for large enough~\fslots  the second term on the RHS of~\eqref{eq: App II lb discrete time}    admits the  following closed-form approximate integral expression
\begin{multline}
 \limK \frac{1}{\tslots\tstep} \logdet{\I + \frac{\Pave\tstep}{\fslots} \cov{\Hv}}  \\ 
 \approx \fslots\fstep\! \inttaunulog{1+\frac{\Pave}{\fslots \fstep}\scaFtn}\label{eq: approx theorem 2}.
\end{multline}
 The proof is complete by replacing~\eqref{eq: approx theorem 2} in~\eqref{eq: App II lb discrete time} and substituting $\bandwidth=\fslots \fstep$.
 
\vspace{-0.9\baselineskip}
\bibliographystyle{IEEEtran}
\bibliography{IEEEabrv,publishers,confs-jrnls,giubib}

\end{document}

%% file: standard-macros.tex
\usepackage{amsmath}
\usepackage{amssymb}

\usepackage{xspace}

\newcommand{\safemath}[2]{\newcommand{#1}{\ensuremath{#2}\xspace}}

\newcommand{\lefto}{\mathopen{}\left}

\newcommand{\takeout}[1]{}

\def\be{\begin{equation}}
\def\ee{\end{equation}}

\def\ba#1\ea{\begin{align*}#1\end{align*}} 
\def\bal#1\eal{\begin{align}#1\end{align}} 
\def\bs#1\es{\begin{split}#1\end{split}} 
\def\bg#1\eg{\begin{gather}#1\end{gather}} 
\def\bi#1\ei{\begin{itemize}#1\end{itemize}}

\DeclareMathOperator{\Prob}{\mathcal{P}}

\newcommand{\circnorm}{\mathcal{CN}} 

\def\det{\qopname\relax o{det}}

\newcommand{\abs}[1]{\lvert#1\rvert}

\newcommand{\distributedas}{\sim}

\newcommand{\conj}[1]{\ensuremath{#1^{\ast}}}  
\newcommand{\tp}[1]{\ensuremath{#1^{T}}}          
\newcommand{\herm}[1]{\ensuremath{#1^{H}}}    

\safemath{\const}{\mathit{const}}

\newcommand{\all}[2]{#1=0,1,\ldots,#2-1}

\newcommand{\setA}{\mathcal{A}}

\newcommand{\setC}{\mathcal{C}}

\newcommand{\bH}{\mathbf{H}}

\newcommand{\bX}{\mathbf{X}}
\newcommand{\bY}{\mathbf{Y}}

\newcommand{\bmh}{\mathbf{h}}

\newcommand{\bmn}{\mathbf{n}}

\newcommand{\bmx}{\mathbf{x}}
\newcommand{\bmy}{\mathbf{y}}
\newcommand{\bmz}{\mathbf{z}}


%% file: defs.tex
\newlength{\dummybflen}
\newcommand{\mathbfcal}[1]{\mathcal{#1}%
\settowidth{\dummybflen}{$\mathcal{#1}$}%
\hspace{-\dummybflen}\hspace{0.08mm}%
\mathcal{#1}%
\hspace{-\dummybflen}\hspace{0.08mm}%
\mathcal{#1}%
\hspace{-\dummybflen}\hspace{0.08mm}%
\mathcal{#1}%
}
\renewcommand{\mathbfcal}[1]{\boldsymbol{\mathcal{#1}}}

\let\time\undefined

\safemath{\time}{t}				
\safemath{\freq}{f}		
\safemath{\delay}{\tau}
\safemath{\doppler}{\nu}
\safemath{\powerDopplerProfile}{S}
\safemath{\nepero}{e}
\safemath{\noiseDen}{N_{0}}
\safemath{\maxDoppler}{\nu_{0}}
\safemath{\maxDelay}{\tau_{0}}
\safemath{\spread}{\Delta_{\opH}}

\safemath{\dtime}{k}					
\safemath{\dfreq}{n}

\safemath{\tstep}{T}					
\safemath{\fstep}{F}

\safemath{\Dtime}{\Delta \time}	                                          
\safemath{\Dfreq}{\Delta \freq}	                                          

\safemath{\fslots}{N}	
\safemath{\tslots}{K}	
\safemath{\tslotsprime}{\tslots^{'}}	

\safemath{\bandwidth}{W}

\safemath{\ppar}{\beta} 			

\safemath{\rateFSKinf}{R_{\text{FSK},\infty}}

\safemath{\cawgninf}{C_{\text{AWGN},\infty}}

\safemath{\alphanew}{\alpha(\bandwidth)}

\newcommand{\sampleTsampleF}{\dtime \tstep,\dfreq \fstep}
\newcommand{\sampleTsampleFprime}{\dtime' \tstep,\dfreq' \fstep}
\safemath{\dfdt}{\dtime,\dfreq}
\safemath{\dfdtprime}{\dtime',\dfreq'}
\safemath{\TF}{\tstep\fstep}

\safemath{\inp}{x}         			
\safemath{\outp}{y}       			
\safemath{\noise}{z}      		
\safemath{\outpnonoise}{r}

\safemath{\inpdiscr}{x}
\safemath{\inpdiscriid}{\inpdiscr_{\text{\iid}}}			
\safemath{\outdiscr}{y}
\safemath{\noisediscr}{z}
\safemath{\chdiscr}{h}

\safemath{\inpdetscal}{x}
\safemath{\inpdetscaltilde}{\tilde{x}}
\safemath{\inpdetvec}{\bmx}
\safemath{\inpdetscaliid}{\inpdetscal_{\text{\iid}}}

\safemath{\X}{\bmx}
\safemath{\Xiid}{\X_{\text{\iid}}}
\safemath{\Y}{\bmy}
\safemath{\Z}{\bmz}
\safemath{\Hv}{\bmh}

\safemath{\Hvoo}{\bH_{0,0}}
\safemath{\Hlinea}{\overline{\bH}}
\safemath{\Hvt}{\widehat{\bH}}
\safemath{\Hvtilde}{\widetilde{\bmh}}
\safemath{\x}{\bmx}
\safemath{\xtilde}{\tilde{\x}}
\safemath{\n}{\bmn}
\safemath{\y}{\bmy}
\safemath{\Ht}{\widehat{H}}
\safemath{\xss}{\mathsf{X}}
\safemath{\Xtilde}{\overline{\X}}
\safemath{\sx}{\lVert \x \rVert ^2}

\safemath{\YY}{\bY}
\safemath{\XX}{\bX}
\safemath{\HH}{\mathbfcal{H}}

\safemath{\opH}{\mathbb{H}}

\safemath{\nks}{\lefto[\dfreq,\dtime\right]}
\safemath{\oo}{\lefto[0,0\right]}

\safemath{\Sig}{\mathsf{\Sigma}}
\safemath{\sigmaH}{\sigma^2_{\opH}}
\safemath{\sigmaZ}{\sigma^2_Z}

\newcommand{\cov}[1]{\mathbf{C}_{#1}}

\newcommand{\scafun}[1]{C_{\opH}\lefto(#1\right)}  
\newcommand{\trafun}[1]{L_{\opH}\lefto(#1\right)}    
\newcommand{\trafunconj}[1]{\conj{L}_{\opH}\lefto(#1\right)}    
\newcommand{\corfun}[1]{R_{\opH}\lefto(#1\right)}   

\newcommand{\ch}[1]{h\lefto(#1\right)}                       

\safemath{\scaFtn}{\scafun{\delay,\doppler}}

\newcommand{\scafuntilde}[1]{\widetilde{C}_{\opH}\lefto(#1\right)}

\safemath{\I}{\mathbf{I}}
\safemath{\Fo}{\mathsf{F}}
\safemath{\zero}{\mathbf{0}}

\newcommand{\media}{\mathbb{E}}

\newcommand{\EV}[2]{\media_{#1}\lefto\lbrace #2 \right\rbrace }

\newcommand{\trace}[1]{\mathrm{Tr}\lefto( #1 \right) }

\newcommand{\diag}[1]{\mathrm{diag}\lefto\lbrace #1 \right\rbrace }
\newcommand{\logdet}[1]{\log \det\lefto(#1\right)}
\newcommand{\determinant}[1]{\mathrm{det}\lefto( #1 \right) }

\safemath{\avP}{\EV{}{\lVert \X \rVert^2}\leq \tslots\Pave\,\tstep}

\safemath{\avPKprime}{\EV{}{\lVert \X \rVert^2}\leq \tslots'\Pave \tstep}

\safemath{\avRho}{\EV{}{\lVert \X \rVert^2} = \tslots \rho\, \tstep}

\safemath{\avalpha}{\EV{}{\lVert \X \rVert^2} = \alpha\tslots\Pave\, \tstep}

\safemath{\avRhoSecond}{\frac{1}{\tslots\rho \tstep}\int \sx \id \probx(\x) = 1 }

\safemath{\avPSecond}{\frac{1}{\tslots\Pave \tstep}\int \sx \id \probx(\x) \leq 1 }

\newcommand{\probprime}{\widetilde{\Prob}_{\!\X}}

\newcommand{\probx}{\Prob_{\!\X}}

\newcommand{\probinpdiscr}{\Prob_{\!\inpdiscr}}

\newcommand{\id}{\mathrm{d}}

\newcommand{\inttaunu}[1]{\int_{\delay}\int_{\doppler} \log\lefto( #1 \right) \id \delay \id \doppler}

\newcommand{\inttaunulog}[1]{\int_{\delay}\int_{\doppler} \log\lefto( #1 \right) \id \delay \id \doppler}

\newcommand{\intnu}[1]{\int_{\nu} \log\lefto( #1 \right) \id \nu}

\newcommand{\intlim}{\int_{-\infty}^{\infty}}

\safemath{\Ppeak}{P_{\mathrm{peak}}}

\safemath{\PTN}{\frac{\Ppeak \tstep}{\fslots}}

\safemath{\Pave}{P}

\safemath{\inpSet}{\mathcal{X}}

\safemath{\inpSetTilde} {\widetilde {\mathcal{X} } }

\safemath{\XpeakTF}{\inpSet_{\fslots\tslots}(\Ppeak)}

\safemath{\XpeakTFo}{\mathcal{X}^{(0)}_{NK \MT}(\Ppeak)}

\safemath{\XpeakTFMT}{\mathcal{X}_{NK \MT}(\Ppeak)}

\safemath{\XpeakMinus}{\XpeakTF \setminus \{\mathbf{0}\}}

\safemath{\XpeakMTMinus}{\XpeakTFMT \setminus \{\mathbf{0}\}}

\safemath{\XpeakSymb}{\widetilde{\mathcal{X}}_{NK}\left(\Ppeak\right)}

\safemath{\XpeakSymbprime}{\widetilde{\mathcal{X}}_{NK'}(\Ppeak)}

\safemath{\peP}{\x \in \XpeakTF}

\safemath{\pePMT}{\x \in \XpeakTFMT}

\safemath{\pePo}{\x_{0} \in \XpeakTFo}

\safemath{\pePsymb}{\x \in \XpeakSymb}

\safemath{\pePsymbKprime}{\x \in \XpeakSymbprime}

\safemath{\limK}{\lim_{ \tslots\rightarrow \infty}}

\safemath{\limKprime}{\lim_{K'\rightarrow \infty}}

\safemath{\limN}{\lim_{\fslots\rightarrow \infty}}

\safemath{\supPeakAve}{\sup_{\probx}}

\safemath{\supPeakAveMT}{\mathop{\mathop{\sup_{\probx}}_{\avP}}_{\pePMT}}

\safemath{\supPeakRho}{\sup_{\probprime}}

\safemath{\supPeakRhoMT}{\mathop{\mathop{\sup_{\probx}}_{\avRho}}_{\pePMT}}

\safemath{\supPeakSymbAve}{\sup_{\probx } }

\safemath{\supPeakSymbAveKprime}{\mathop{\mathop{\sup_{\probx }}_{\avPKprime}}_{\pePsymbKprime}}

\safemath{\supPeakSymb}{\sup_{\pePsymb}}

\safemath{\supPeakSymbAveSecond}{\mathop{\mathop{\sup_{\probx}}_{\avPSecond}}_{\pePsymb}}

\safemath{\supAve}{\mathop{\sup_{\probx}}_{\avP}}

\safemath{\supalpha}{\sup_{0\leq \alpha \leq 1}}

\safemath{\supPeak}{\sup_{\peP}}

\safemath{\supCov}{\mathop{\sup_{\cov{\X}}}_{\trace{\cov{\X}}\leq KPT}}

\safemath{\supCovMT}{\mathop{\sup_{\cov{\X\oo}}}_{\trace{\cov{\X\oo}}\leq KPT}}

\safemath{\infPeakRho}{\inf_{\probprime}}

\safemath{\infPeakRhoMT}{\mathop{\mathop{\inf_{\probx}}_{\avRho}}_{\pePMT}}

\safemath{\infRhoSecond}{\mathop{\inf_{\probx}}_{\avRhoSecond}}

\safemath{\infPeak}{\inf_{\peP}}

\safemath{\infPeako}{\inf_{\pePo}}

\safemath{\infPeakSymb}{\inf_{\pePsymb}}

\safemath{\infPeakSet}{\inf_{\x}}

\safemath{\infPeakSetTilde}{\inf_{\xtilde}}

\safemath{\infPeakSymbSet}{ \mathop{ \mathop{ \mathop{\inf_{\x \in \XpeakSymb}}_{|x_{n,k}|^{2} \in 
\left\{ 0,\frac{\Ppeak T}{N} \right \} } }_{ \all{n}{N} } }_{ \all{k}{K} } }

\safemath{\infPeakRhoSec}{\mathop{\mathop{\inf_{\probx}}_{\avRhoSecond}}_{\peP}}

\safemath{\UBone}{\mathrm{UB}_1}

\safemath{\UBtwo}{\mathrm{UB}_2}

\safemath{\LB}{\mathrm{LB}}

\safemath{\LBtwo}{\mathrm{LB}_2}

\safemath{\xtwonk}{\lvert x_{n,k} \rvert^{2}}

\safemath{\nk}{(n,k)}

\safemath{\onebold}{\mathbf{1}\lefto\{  \setA  \right\}}

\safemath{\setAhat}{\setC}

\newtheorem{Theor}{Theorem}

\newcommand{\esp}[2]{e^{#1j2\pi #2}}

\safemath{\MT}{M_{T}}

\safemath{\MR}{M_{R}}

\safemath{\infcapacity}{C_{\infty}}

\newcommand{\closedform}{closed-form\xspace}

\safemath{\logtwo}{\log_{2}}
\let\to\undefined
\safemath{\to}{\rightarrow}
\newcommand{\iid}{i.i.d.\@\xspace}